\documentstyle[12pt]{article}
\newcommand{\be}{\begin{equation}}
\newcommand{\ee}{\end{equation}}
\newcommand{\ba}{\begin{eqnarray}}
\newcommand{\ea}{\end{eqnarray}}

\begin{document}
\begin{center}       
{\bf\Large
{Killing-Yano Symmetry for a Class of Spacetimes Admitting Parallel Null
1-Planes}}  
\end{center}
\begin{center}
D. Baleanu\footnote
{On leave of absence on
Institute of Space Sciences, P.O.Box, MG-23, R 76900,\\
Magurele-Bucharest, Romania,
E-mail:baleanu@venus.nipne.ro}
\end{center}
\begin{center}
 Department of Mathematics and Computer Science,\\
Faculty of Arts and Sciences, {\c C}ankaya University,\\
 Ankara, 06350 Balgat, Turkey
\end{center}    
\begin{center}
and
\end{center}

\begin{center}
{S. Ba{\c s}kal \footnote{E-mail:
baskal@newton.physics.metu.edu.tr}}
\end{center}

\begin{center}
Department of Physics, Middle East Technical University,
06531 Ankara, Turkey  
\end{center}

\begin{abstract}
A possible generalization of plane fronted waves with parallel
rays (gpp-wave) fall into a more general class of metrics admitting 
parallel null 1-planes.
For gpp-wave metric, the zero-curvature condition is given,
the Killing-Yano tensors of order two and three are found and 
the corresponding Killing tensors are constructed.  
Henceforth, the compatibility between geometric duality and 
non-generic symmetries is presented.
\end{abstract}

\section{Introduction}
\noindent
The plane fronted waves with parallel rays (pp-wave) \cite{kundt} is one 
of those certain metrics that receive much attention in the literature,
as well as its generalized forms which were studied from various 
aspects \cite{pod98,baskal99}.  On the other hand, Walker's metric 
\cite{walker} is also interesting in its own right.  For instance, in four 
dimensions, for each field of null 1-plane, it is possible to associate a 
"classical" massless particle, with its four momentum vector to be the 
basis of the 1-planes.  Furthermore, the recurrence vector of such a plane 
was interpreted to represent the electromagnetic potential \cite{oktem73}.   
It has been shown that the canonical form of Walker's metric can be
brought into a simpler form, by an appropriate coordinate transformation
\cite{akcag78} and the resulting form can always be diagonalized within 
the components related to the space part of the metric.
Kundt's metric, and some of its generalizations, as will be noted in
this paper, fall into a subclass of Walker's metric.

Recently, the generic and especially non-generic symmetries
\cite{gibbons, holten, visinescu, tanimoto, ali00}
of the spinning space \cite{bardu76,brink76,berez77,casal76}
have been investigated by several authors. 
The Killing-Yano (KY) tensors  of rank two or higher order \cite{yano, 
tachibana, collinson, kramer, rudinger, hall} play an important role 
for new non-generic symmetries as well as the generalized KY tensors 
\cite{laur}, which are deeply connected to generalized Runge-Lenz symmetry. 
Also recently, it was discovered that there is a connection between Lax 
tensors  \cite{rosquist,bb1} and KY tensors of order three.  

Within the same context of the Killing and KY tensors, the notion of 
geometric duality was defined and applied on Taub-NUT and Kerr-Newman 
spacetimes \cite{holten96}. 
The direct way to construct the dual metrics is to calculate the Killing
tensors \cite{bb2}.  An alternative way is to find the KY tensor and contract 
two of them to find Killing tensors \cite{holten96, visinescu1}.  
However, finding an explicit spacetime with a physical significance
admitting higher rank KY tensors is not an easy task, not only because
of the complexity of the calculations, but also earlier it was found that
not all metrics admit KY tensors of higher order \cite{dub99}.
Furthermore, it is worthwhile to compare the dual metrics obtained from 
the Killing tensors that are found by solving their defining equations 
\cite{bb2}, with those to be investigated in this paper. 
Henceforth, we believe to have a fuller understanding of geometric duality.

The plan of our paper is the following: In Sec. 2 we give a brief 
review of non-generic supercharges and geometric duality.
In Sec. 3 we give the generalized form of the pp-wave metric. 
In Sec. 4 we present the KY and the Killing tensors.  The dual 
metrics will also be given if they exist.  Section 5
is devoted to concluding remarks.  Finally, in the Appendix, we have 
presented the pure radiation condition and the Petrov classification  
of the background metric admitting KY tensors.

\section{Killing-Yano tensors and the Dual Metrics}
\subsection{Killing-Yano tensors and non-generic supercharges}
\noindent
An $n^{th}$ rank Killing-Yano tensor $f_{\nu_{1}\nu_{2}\cdots\nu_{n}}$,
is an antisymmetric tensor fulfilling the following equations:
\begin{equation}
f_{\nu_{1}\nu_{2}\cdots(\nu_{n};\lambda)}\,=\,0,
\label{kye}
\end{equation}
where semicolon denotes the covariant derivative.

The spinning particle model was constructed to be
supersymmetric \cite{bardu76,brink76,berez77,casal76},  
therefore independent of the form of the 
metric there are always four independent generic symmetries
\cite{gibbons, ali00}. 
The existence of Killing-Yano tensors of valence $n$ are related to 
non-generic supersymmetries described by the supercharge
$ Q_{f}=f_{\nu_{1}\nu_{2}\cdots\nu_{r}}\Pi^{\nu_{1}}\psi^{\nu_{2}}\cdots
\psi^{\nu_{r}} $ which is a superinvariant: $\{Q_{0}, Q_{f}\}=0$,
where $Q_{0}=\Pi_{\mu}\psi^{\mu}$.
Here, $\Pi_{\mu}$ is the covariant momenta and $\psi^{\mu}$ are odd
Grassmann variables.       
The KY equation and the Jacobi identities guarantee that it is
also a constant of motion: $\{Q_{f},H\}=0$, with
$ H=\frac{1}{2}g^{\mu\nu}\Pi_{\mu}\Pi_{\nu},$
and with the appropriate definitions of the brackets.

\subsection{Killing tensors and dual spacetimes}
\noindent
A Killing tensor of valence two is defined through the equation
\begin{equation}
K_{(\mu\nu;\alpha)}=0.
\label{kte}
\end{equation}
Killing-Yano tensors of any valence can be considered as the square root 
of the Killing tensors of valence two in the sense that, their appropriate
contractions yield
\begin{equation}
K_{\mu\nu}=g^{\alpha\beta}f_{\mu\alpha}f_{\beta\nu}
\label{kfky1}
\end{equation}
or for valence three it can be written as
\begin{equation}
K_{\mu\nu}=g^{\alpha\delta}g^{\beta\gamma}
f_{\mu\alpha\beta}f_{\gamma\delta\nu}.
\label{kfky2}
\end{equation}

It has been shown in detail in reference \cite{holten96} that $K^{\mu\nu}$
and $g^{\mu\nu}$ are reciprocally the contravariant components of the
Killing tensors with respect to each other.
If $K^{\mu\nu}$ is non-degenerate, then through the relation
\begin{equation}
K^{\mu \alpha}k_{\alpha \nu}=\delta^{\mu}\,_{\nu},
\label{dual}
\end{equation}
the second rank non-degenerate tensor $k_{\mu\nu}$, 
can be viewed as the metric on the "dual" space.
Furthermore, the notion of geometric duality extends to that of 
phase space.  The constant of motion 
$K=\frac{1}{2}K^{\mu\nu}p_{\mu}p_{\nu}$,
generates symmetry transformations on the phase space linear
in momentum: $\{ x^{\mu},K \}=K^{\mu \nu}p_{\nu}$.
The Poisson brackets satisfy $\{ H,K \}=0$, 
where $H=\frac{1}{2}g^{\mu\nu}p_{\mu}p_{\nu}$.  Thus, in the phase 
space there is a reciprocal model with constant of motion $H$ and 
the Hamiltonian $K$. 

\section{Subclasses of Parallel Null 1-Planes}
\noindent
The well-known pp-wave metric is given in the form \cite{kundt}:
\begin{equation}
ds^{2}=2\,dv\,du + dx^{2} + dy^{2} +
H(x,y,u)\,du^{2}.
\end{equation}
One possible generalization of this metric
can be written as:
\begin{equation}
ds^{2}=2\,dv\,du + A(x,y,u)\left[dx^{2} + dy^{2}\right] +
H(v,x,y,u)\,du^{2}
\label{om}
\end{equation} 
and, hereafter we shall refer it as the gpp-wave metric.
With such a generalization it falls into a class of spacetimes
admitting a parallel field of null 1-planes.  This latter, consists 
in a recurrent field of null vectors.  If $l_{\mu}$ is a basis for 
the plane we have:
\begin{equation}
\nabla_{\nu} \, l_{\mu} = \kappa_{\nu} l_{\mu},
\qquad 
l_{\mu}l^{\mu}=0, \quad l_{\mu} \ne 0
\label{pn1p}
\end{equation}
where $\kappa_{\mu}$ is the recurrence vector of the plane. 
It can be seen that such a vector field is geodesic and non-rotating.
>From (\ref{pn1p}) and the Ricci identity we have
\begin{equation}
l^{\nu}R_{\mu \nu \, \alpha \beta}=l_{\mu}t_{\alpha \beta},
\label{biv}
\end{equation}
where $ t_{\alpha \beta}=\partial_{\alpha}\kappa_{\beta} 
- \partial_{\beta}\kappa_{\alpha}$. Also from
(\ref{pn1p}) one can observe that the principal null 
vector $l_{\mu}=\delta_{\mu}\,^{4}=\partial_{\mu} u$, is 
hypersurface-orthogonal and the recurrence vector for the PN1P is 
$\kappa_{\mu}=-\Gamma^{4}\,_{\mu 4}$.

The canonical form of the metric admitting parallel field of
null 1-planes is given by Walker \cite{walker}, and that also
can be brought into a simpler form, by an appropriate the coordinate
transformation \cite{akcag78}.  Then the resulting form can always be
diagonalized within the metric functions $g_{22},g_{33}$ and $g_{23}$. 
As such, the simplified form becomes:
\begin{equation}
ds^{2}=2\,dv\,du + A(x,y,u)\, dx^{2} + B(x,y,u)\,dy^{2} +
H(v,x,y,u)\,du^{2}
\label{mm}
\end{equation}                        
where $A(x,y,u),B(x,y,u)$ and $H(v,x,y,u)$ are functions of their
arguments.  This metric reduces to (\ref{om}), when $B=A$.  

In the following analysis, the scalar curvature for (\ref{om}) 
will prove to be important.  It is calculated to be:   
\begin{equation}
R=\frac{1}{A^{3}} [ A_{,x}^{2}+A_{,y}^{2}-A(A_{,xx}+A_{,yy}) ].
\label{r}
\end{equation}

\section{Solutions to the KY equations, Killing tensors and the dual metrics}
\noindent
In this section the KY equations for rank two and three will be 
investigated. There are 24 independent equations, to
be solved for the six independent components of the
Killing-Yano tensor of rank two
and 15 independent equations corresponds  for four independent components
of the KY tensor of rank three respectively. 
We will look for the subclasses of the gpp-wave metric 
admitting KY tensors with each and every number of surviving components
ranging from one to six for rank two, and from one to four for
rank three.  We will eliminate all flat background solutions.
Having in mind the importance of the manifolds with scalar curvature 
\cite{havas} we will investigate the $R=0$ and $R \neq 0$ cases 
separately.  Using the method of separation of variables,  
the solution to the $R=0$ equation in (\ref{r}) is obtained as:
\begin{equation}
A(x,y,u)=a(u) \exp
\{ {c_{1}\over 2}[(y-c_{2})^{2}-(x-c_{2})^{2}] \},
 \label{rsz}
\end{equation}
where $c_{1}$ and $c_{2}$ are constants and $a$ depends only on u.        
The $g_{44}$ component of the metric plays an important role in solving 
the KY equations.  The consistency conditions of KY equations leads us to 
the form of $H(x,y,u,v)$  as 
\begin{equation}
H(x,y,u)=v\,h1(x,y,u) +h2(x,y,u).
\label{eych}
\end{equation}

\subsection{Second rank KY tensors}
\noindent
In the following we will present the KY tensor $f_{\mu\nu}$,
the corresponding Killing tensor $K_{\mu\nu}$ obtained by using
(\ref{kfky1}), the  dual metric $k_{\mu\nu}$ if it exists, and the 
form of the metric, in that order.  All solutions are for $R=0$, unless
otherwise stated.\\  
{\it One component solutions:}\\
${\bf a})$
\begin{equation}
f_{14}=c.
\end{equation}
Then
\begin{equation}
K_{14}=c^{2}, \qquad K_{44}=c^{2}H(u,v).
\end{equation}
Here, $c$ is an arbitrary constant. The metric function $A$ is independent 
of $u$, and 
 $H(v,u)=v\,h1(u)+h2(u)$.

\noindent
${\bf b})$
\begin{equation}
 f_{23}=c\,A(x,y),
\end{equation}
and
\begin{equation}
K_{22}=K_{33}=-c^{2}A(x,y).
\end{equation}
Here, $c$ is an arbitrary constant and $A$ and $H$ are as above. 

\noindent
${\bf c})$  
\begin{equation}
f_{24}=A(u)^{\frac{1}{2}}\,Q(u),
\end{equation}
and we obtain
\begin{equation}
K_{44}=-Q(u)^{2},
\end{equation}
where
\begin{equation}
Q(u)=\exp\left[\frac{-1}{2} \int h1(u)\,du \right].
\label{qu}
\end{equation}
Now, $A$ is a function of only $u$ and $H(v,x,y,u)=v\,h1(u)+h2(x,y,u)$.

\noindent
${\bf d})$ For $f_{34}$ the solution is as above.

\noindent
${\bf e})$ For $R \neq 0,$ we have
\begin{equation}
f_{23}=g(u)^{3/2}s(x,y). 
\end{equation}
Then
\begin{equation}
K_{22}=K_{33}=-g(u)^{2}s(x,y),
\end{equation}
with $A(x,y,u)=g(u)s(x,y)$, where $s(x,y)$ is an arbitrary function
of its arguments and $ H(v,u)=v\,h1(u)+h2(u)$.  This is the only solution 
for a second rank KY tensor with a non-zero scalar curvature.

\noindent
{\it Two component solutions:}\\[2mm]
${\bf a})$ 
\begin{equation}
f_{14}=c_{3}, \qquad f_{23}=c_{4}\,A(x,y),
\end{equation}
so
\begin{equation}
K_{14}= c_{3}^{2}, \qquad
K_{22}=K_{33}=-c_{4}^{2}A(x,y),\qquad
K_{44}=c_{3}^{2}H(v,u)
\end{equation}
and
\begin{equation}
k_{14}=1/ c_{3}^{2}, \qquad
k_{22}=k_{33}= -A(x,y)/c_{4}^{2}\qquad
k_{44}=H(v,u)/c_{3}^{2}. 
\end{equation}
Here, $c_{3}$ and $c_{4}$ are constants, $A$ is independent 
of $u$, and $H(v,u)=v\,h1(u)+h2(u)$.  This is the only
case where we have the compatibility between the non-generic 
symmetries and the dual space.

\noindent
{\bf b)} 
\begin{equation}
f_{23}=A^{3\over 2}(u),  \qquad f_{24}=A(u)^{\frac{1}{2}}P(u).
\end{equation}
We find
\begin{equation}
K_{22}=K_{33}=-A(u)^{2}, \qquad
K_{34}=-A(u)\,P(u),\qquad
K_{44}=-P(u)^{2},
\end{equation}
where 
\begin{equation}
P(u)=Q(u)
\left[
\frac{\epsilon}{2} 
 \int \left\{ h3(u)\,Q(u)^{-1} \right\} \,du + c \right].
\label{pu}
\end{equation}
with $Q(u)$ is as in (\ref{qu}), $c$ is an integration constant and
here $\epsilon=-1$.  The metric function $A$ depends only on
$u$ and $H(v,y,u)=v h1(u) + y h3(u)$.

\noindent
{\bf c)} 
\begin{equation}
f_{23}=A(u)^{3 \over 2}, \qquad
f_{34}=A(u)^{1 \over 2}\,P(u),
\end{equation}
and
\begin{equation}
K_{22}=K_{33}=-A(u)^{2}, \qquad
K_{34}=-A(u)P(u),\qquad
K_{44}=-P(u)^{2},
\end{equation}
where $P(u)$ is as in (\ref{pu}), with $\epsilon=1$, $A$ is a function of 
$u$ and  $H=v h1(u) + x\,h3(u)$.

\noindent
{\bf d)} 
\begin{equation}
f_{24}=f_{34}=A(u)^{\frac{1}{2}}\, Q(u), 
\end{equation}
so
\begin{equation}
K_{44}=-2\,Q(u)^{2}.
\end{equation}
Here, $A$ depends only on $u$, and $H(v,x,y,u)=v\,h1(u)+h2(x,y,u)$.

\noindent
{\it Three component solutions:}\\
{\bf a})~~We have $f_{23},f_{24},f_{34}$ as surviving components.
>From the integrability conditions two cases can be distinguished
in regards to the metric function $h2$.\\
Case {\bf a1)}~~h2 is only a function of u. 
\begin{equation}
f_{23}=A^{3 \over 2}(u)r(u), \qquad f_{24}=-y s(u),\qquad f_{34}=x s(u).
\label{fonea}
\end{equation}
Then
\begin{equation}
\begin{array}{l}
K_{22} = K_{33} = -A(u)^{2}r(u)^{2}, \qquad
K_{24} = x\,A(u)^{\frac{1}{2}}r(u)s(u), \\[2mm]
K_{34} = y\, A(u)^{\frac{1}{2}}r(u)s(u), \qquad
K_{44} = -s(u)^{2}(x^{2}+y^{2})/A(u),
\label{konea}
\end{array}
\end{equation}
where
\begin{equation}
r(u)=\int A(u)^{-1} Q(u)\, du, \qquad
s(u)=A(u)^{\frac{1}{2}}\,Q(u) 
\label{onea}
\end{equation}
with $Q(u)$ as in (\ref{qu}), $A$ depending only on $u$ and 
$H(u,v)=v h1(u)+h2(u)$.\\
\noindent
Case {\bf a2})~ h2 is quadratic in $x$ and $y$.
The components are the same as (\ref{fonea}) and (\ref{konea}),
with $H(u,x,y,v)=v h1(u)+h3(u)(x^{2}+y^{2})$.
The functions $r(u),\, s(u)$ are as in (\ref{onea})
and $A(u)$ is subject to the solutions of 
\begin{equation}
\frac{A(u)_{,u}}{A(u)} - h1(u)
+2A^{3 \over 2}(u)h3(u)\frac{r(u)}{s(u)}=0.
\label{sol}
\end{equation}

\noindent
{\bf b}) 
\begin{equation}
f_{14}=c, \qquad  f_{24}=-c\, x A(u)_{,u} /2, \qquad 
f_{34}=-c\, y\, A(u)_{,u}/2.
\end{equation}
We have
\begin{equation}
\begin{array}{l}
K_{14}= c^{2}, \qquad 
K_{24}= -\frac{1}{2}c^{2}x\,A(u)_{,u},\qquad 
K_{34}= -\frac{1}{2}c^{2}y\,A(u)_{,u},\\[2mm]
K_{44}= c^{2}\left[H(v,x,y,u)-(x^{2}+y^{2})A(u)_{,u}^{2}/4A(u)\right].
\end{array}
\end{equation}
In this case $A$ depends on $u$, $H(v,x,y,u)=v h1(u)+ h3(u)(x^2+y^2)$,
and $h1,h3,A$ are related as
\begin{equation}
2\, A(u)A(u)_{,uu} - A(u)^{2}_{,u} + A(u)_{,u}A(u)h1(u) + 4 A(u) h3(u)=0.
\end{equation}

\subsection{Third rank KY tensors}
{\it Two component solution:}\\
{\bf a})
\begin{equation}
f_{124}=A(u)^{\frac{1}{2}}, 
\qquad f_{234}=\frac{1}{3}\,y\,A(u)^{\frac{3}{2}}\,_{,u}.
\end{equation}
Then
\begin{equation}
\begin{array}{l}
K_{14}= 2, \qquad 
K_{22}= 2\, A(u), \qquad
K_{34}= -y\,A(u)_{,u}\,,\\[2mm]
K_{44}=2\,H(v,x,u)-y^{2}A(u)_{,u}^{2}/2A(u). 
\end{array}
\end{equation}
The metric functions are $A(u)$ and $H=vh_{1}(u)+h_{2}(x,u)$, with
$A(u)$ and $h1(u)$ related as:
\begin{equation}
2\,A(u)\,A(u)_{,uu}-A(u)_{,u}^{2}+A(u)\,A(u)_{,u}h1(u)=0.
\label{te1}
\end{equation}
Such a spacetime is pure radiative, and is not conformally flat.

\noindent
{\bf b})
\begin{equation}
f_{134}=A(u)^{\frac{1}{2}}, \qquad
f_{234}=-\frac{1}{3}\,x\,A(u)^{\frac{3}{2}}\,_{,u}.
\end{equation}
We obtain
\begin{equation}
\begin{array}{l}
K_{14}= 2, \qquad 
K_{24}= -x A(u)_{,u}\,, \qquad
K_{33}= 2\,A(u),\\[2mm]
K_{44}=2\,H(v,y,u)-x^{2}A(u)_{,u}^{2}/2A(u).
\end{array}
\end{equation}
The metric function $A$ depends on $u$, $H=vh_{1}(u)+h_{2}(y,u)$, with
$A(u)$ and $h1(u)$ are relates as in (\ref{te1}).
Such a spacetime is pure radiative, and is not conformally flat.

\noindent
{\bf c}) If  $R \neq 0$ we have the following solution:
\begin{equation}
f_{123}=u\,A(x,y,u), \qquad f_{234}=r(u)A(x,y,u).
\end{equation}
Then
\begin{equation}
\begin{array}{l}
K_{11}= -2\,u^{2}, \qquad
K_{14}= -2\,u \,r(u), \\[2mm]
K_{22} = K_{33}= 2\,u\,A(x,y,u)\left[u\,H(u,v)-2\,r(u)\right],\qquad
K_{44}=-2r(u)^{2}
\end{array}
\end{equation}
We have $A(x,u,y)=u\,s(x,y)$, where $s(x,y)$ is an 
arbitrary function of its arguments and $H=\frac{v}{u}+h2(u)$.
The functions $r(u)$ and $h2(u)$ are related as:
\begin{equation}
2u\,r(u)_{,u}+r(u)-u^{2}h2(u)_{,u}-u\,h2(u)=0.
\end{equation}
This metric is not conformally flat.

\section{Concluding remarks}
In this paper, we have presented a possible generalization
of a pp-wave metric which manifestly proved to fall
into a subclass of metric admitting parallel null 1-planes.
The existence of the KY tensors of order two 
and three for this gpp-wave metric has been investigated. 
Our work was divided in two parts with respect to when the 
scalar curvature is zero or not because those cases arose
from the consistency conditions of KY equations.  A subclass of 
this metric admitting null scalar curvature was found. 
The existence of third rank KY tensors suggests that some subclasses 
has new supercharges.
We observe that Killing tensors generated by KY tensors of order three 
are not invertible.  Nevertheless, we have found a subclass for which the 
geometric duality and non-generic symmetries are compatible. 
We have also presented the Petrov types and the pure radiation condition.
The results obtained in this paper strongly encourage us to 
investigate the existence of the generalized KY tensors recently introduced 
by Collinson.  

\section{Acknowledgments}
We would like to thank to F. {\"O}ktem for valuable discussions and
L. Howarth for sending us her thesis.

\section*{Appendix}
 
If $H$ is as in (\ref{eych}) and in general $R \neq 0$, then 
the Einstein tensor for (\ref{om}) can be found to be of the form:
\begin{equation}
G_{\mu\nu}=l_{\mu}k_{\nu}+l_{\nu}k_{\mu}
\end{equation} 
where $k_{\mu}$ are:
\begin{eqnarray}
k_{1} &=& \frac{1}{2A^{3}}
\left[ A\,(A_{,xx}+A_{,yy})-(A_{,x}^{2}+A_{,y}^{2}) \right],\\
k_{2} & = & \frac{-1}{2A^{2}}(A\,A_{,xu}-A_{,u}A_{,x} - A^{2}h1_{,x}),\\
k_{3} & = & \frac{-1}{2A^{2}}(A\,A_{,yu}-A_{,u}A_{,y} - A^{2}h1_{,y}),\\
k_{4} & = & \frac{-1}{4A^{3}}
\{ H\, \left[ A(A_{,xx}+A_{,yy})-(A_{,x}^{2} + A_{,y}^{2}) \right]
 \\\nonumber
& & 
-A\,\left[ 2\,A\,A_{,uu}-A_{,u}^{2}+A\,A_{,u}h1
+A\,(H_{,xx} + H_{,yy}) \right] \}\, .
\end{eqnarray}
One can observe that, when $h1$ is independent of $x$ and $y$ and $R=0$ we 
have pure radiation; i.e., $G_{\mu\nu}=\rho(x^{\sigma}) l_{\mu}l_{\nu}$,
where $\rho(x^{\sigma})$ is the energy density.
          
It is worthwhile to study the algebraic properties of the conformal Weyl
tensor with respect to the principal null vector $ l_{\mu} $.
With $ l^{\mu}=\delta^{\mu}\,_{1} $ the following
characterization for different Petrov types (PT) are obtained:
\begin{equation}
\begin{array}{llll}
\mbox{For} \qquad R=0 :&  & & \\
H=vh1(u)+h2(x,y,u), & \Leftrightarrow&  
\begin{array}{l}
C_{\mu\nu\alpha\beta}\neq 0,\\
l^{\sigma} C_{\mu\sigma\alpha\beta}= 0   
\end{array}   
                                & \Leftrightarrow\;\;\mbox{type N}.\\[10mm]
H=vh1(x,y,u)+h2(x,y,u),& \Leftrightarrow & 
\begin{array}{l}
l^{\sigma} C_{\mu\sigma\alpha\beta}\neq 0, \\
l^{\sigma}l_{[\mu} C_{\nu]\sigma\alpha\beta}= 0 
\end{array}   
                                & \Leftrightarrow\;\;\mbox{type III}.\\[1cm]
\mbox{For}\qquad R\neq 0 :&  & & \\
H=vh1(x,y,u)+h2(x,y,u),  & \Leftrightarrow & 
\begin{array}{l}
l^{\sigma}l_{[\mu} C_{\nu]\sigma\alpha\beta}\neq 0,\\    
              l^{\sigma}l^{\rho}l_{[\mu} C_{\nu]\rho\sigma\beta}= 0
\end{array}    
                               & \Leftrightarrow\;\;\mbox{type II or D}.

\end{array}
\end{equation}
If $R=0$, $h1$ is depending only on $u$ and $h2$ is
depending on $u,x,y$ we present the equations for the Weyl tensor to be 
zero:      
\begin{eqnarray}
h2_{,xx} - h2_{,yy} -c_{1}c_{2}(h2_{,x}+h2_{,y})
+c_{1}(x\,h2_{,x} + y\, h2_{,y})=0, \\
2\,h2_{,xy} + c_{1}c_{2}\,(h2_{,x} - h2_{,y}))
- c_{1}(y\,h2_{,x}+ x\,h2_{,y})=0, \\
h2_{,xx} - h2_{,yy} - c_{1}((c_{2} - x)\,h2_{,x} 
- (c_{2} - y)\,h2_{,y})=0.
\label{weyl1}
\end{eqnarray}
Obviously if $h2$ depends only on $u$ the above equations are fulfilled. 
 
Earlier, Petrov types for PN1P have been investigated, in terms of a second 
rank tensor involving the Ricci tensor and a bivector \cite{oktem73}. 
It can be observed that, when the metric function $H$ 
is as in (\ref{eych}), then $t_{\alpha\beta}$ in (\ref{biv}) vanishes. 
Petrov types of spacetimes $v^{\mu} R_{\mu\nu\alpha\beta}=0$, 
where $v^{\mu}$ is not necessarily null, have also been investigated
from a more general point of view \cite{mcintosh}.  
Here, we have made a further analysis on PT to give 
the explicit forms of the metric functions.

\end{document}